\documentclass[aps,prl,twocolumn,superscriptaddress]{revtex4-1}
\usepackage{amsmath,amssymb,graphicx,natbib,aas_macros}

\newcommand{\ie}{\textit{i.e.}}
\newcommand{\eg}{\textit{e.g.}}

\newcommand{\e}{\ensuremath {\epsilon}}
\newcommand{\VEV}[1]{\ensuremath{\left\langle #1 \right\rangle}}
\newcommand{\Orderof}[1]{{\ensuremath{\mathcal{O}\!\left(#1\right)}}}
\newcommand{\sigmavof}[1]{\ensuremath{\langle \sigma_{{#1}}v \rangle}}

\newcommand{\deut}{\ensuremath{\mathrm{D}}}
\newcommand{\trit}{\ensuremath{\mathrm{T}}}

\newcommand{\Fe}{\ensuremath{\mathrm{Fe}}}
\newcommand{\Hyd}{\ensuremath{\mathrm{H}}}

\newcommand{\bel}{\ensuremath{{}^{11}\mathrm{B}}}
\newcommand{\hef}{\ensuremath{{}^4\mathrm{He}}}
\newcommand{\het}{\ensuremath{{}^3\mathrm{He}}}
\newcommand{\lisx}{\ensuremath{{}^6\mathrm{Li}}}
\newcommand{\lisv}{\ensuremath{{}^7\mathrm{Li}}}

\newcommand{\bes}{\ensuremath{{}^7\mathrm{Be}}}
\newcommand{\beet}{\ensuremath{{}^8\mathrm{Be}}}
\newcommand{\ben}{\ensuremath{{}^9\mathrm{Be}}}
\newcommand{\beten}{\ensuremath{{}^{10}\mathrm{Be}}}
\newcommand{\bt}{\ensuremath{{}^{10}\mathrm{B}}}
\newcommand{\hes}{\ensuremath{{}^6\mathrm{He}}}

\newcommand{\keV}{\ensuremath{\mathrm{keV}}}
\newcommand{\MeV}{\ensuremath{\mathrm{MeV}}}
\newcommand{\GeV}{\ensuremath{\mathrm{GeV}}}
\newcommand{\TeV}{\ensuremath{\mathrm{TeV}}}
\renewcommand{\sec}{\ensuremath{\mathrm{s}}}

\newcommand{\mbarn}{\ensuremath{\mathrm{mb}}}
\newcommand{\barn}{\ensuremath{\mathrm{b}}}

\newcommand{\K}{\ensuremath{\mathrm{K}}}

\begin{document} 

\title{Primordial beryllium as a big bang calorimeter}

\author{Maxim Pospelov}
\email{pospelov@uvic.ca}
\affiliation{Perimeter Institute for Theoretical Physics, Waterloo,
ON, N2L 2Y5, Canada}
\affiliation{Department of Physics and Astronomy, University of
  Victoria, Victoria, BC, V8P 1A1, Canada}

\author{Josef Pradler}
\email{jpradler@perimeterinstitute.ca}
\affiliation{Perimeter Institute for Theoretical Physics, Waterloo,
ON, N2L 2Y5, Canada}

\begin{abstract}
  Many models of new physics including variants of supersymmetry
  predict metastable long-lived particles that can decay during or
  after primordial nucleosynthesis, releasing significant amounts of
  non-thermal energy. The hadronic energy injection in these decays
  leads to the formation of \ben\ via the chain of non-equilibrium
  transformations: ${\rm Energy}_h\to {\rm T},\het \to \hes,\, \lisx
  \to \ben$.  We calculate the efficiency of this transformation and
  show that if the injection happens at cosmic times of a few hours,
  the release of $\Orderof{10\,\MeV}$ per baryon can be sufficient for
  obtaining a sizable \ben\ abundance.  The absence of a
  plateau-structure in the $\ben/\Hyd$ abundance down to a
  $\Orderof{10^{-14}}$ level allows one to use beryllium as a robust
  constraint on new physics models with decaying or annihilating
  particles.
\end{abstract}

\maketitle

The rare light elements lithium, beryllium, and boron play a pivotal
role in observational cosmology. Their abundance pattern is key in our
understanding of stellar structure and galactic evolution. It may also
probe the very beginning of nuclear chemistry, known as big bang
nucleosynthesis (BBN).
Standard BBN (SBBN) theory has become what could be called a
parameter-free theory. With the cosmic microwave background
determination of the baryon asymmetry, $\eta_{b}\simeq 6.23 \times
10^{-10}$~\cite{Dunkley:2008ie}, SBBN solely relies on Standard Model
physics embedded in a Friedman-Robertson-Walker Universe.  An overall
concordance between the SBBN predictions and the observationally
inferred primordial abundances of \deut\ and \hef\ has put BBN on firm
footing and made its framework an invaluable toolbox for testing
models of new physics~(see,
\eg~\cite{doi:10.1146/annurev.nucl.012809.104521}).
Whereas Li, Be, and B are mainly produced in galactic cosmic rays
\cite{1970Natur.226..727R,1971A&A....15..337M}, the only stable
isotope to reach out to a primordial fraction at the ppb level is
\lisv.  The SBBN yields of Be and B are
negligible~\cite{Thomas:1992tq} and, to date, the only known scenarios
which may produce Be in interesting amounts are inhomogeneous
nucleosynthesis, see \eg~\cite{Jedamzik:2001qc}, and the catalysis of
nuclear reactions by electromagnetically or strongly interacting relic
particles~\cite{Pospelov:2007js,Pospelov:2008ta,Kusakabe:2009jt}.

The purpose of this paper is to show that non-equilibrium processes
induced by the decay or annihi\-lation of a relic particle species $X$
after BBN can lead to the production of primordial \ben/\Hyd\ at the
$\mathcal{O}(10^{-14})$ level and above, which makes these scena\-rios
testable via atmospheric measurements of \ben\ in extremely
metal-deficient stars (MDSs).  We identify the following prospective
path to \ben:
\begin{align}
\label{eq:chain}
\begin{array}{ccccccc} {\rm Energy}_h{\to} & \trit,\, \het & \to &
  \hes,\, \lisx &\to &\ben ,
  \end{array}
\end{align}
where Energy$_{h}$ refers to the injection of energy via hadronic
channels.  The chain of transformations is initiated by the spallation
of \hef\ due to ``primary'' nucleons and followed by repeated
reactions on ambient $\alpha$-particles.  It is well known that such
non-thermal cascades---truncated at $A=6$---can be an efficient source
of ``secondary'' \lisx~\cite{Dimopoulos:1987fz,Jedamzik:1999di}, as it
bypasses the extremely inefficient SBBN channel
D($\alpha,\gamma$)\lisx. Here we show that the exoergic reaction
\begin{align}
  \label{eq:rct-hes}
\hes+\hef  & \to \ben + n,  \quad Q = 0.60\ \MeV
\end{align}
is particularly capable of going one step further and bridging the
$A=8$ divide.  As we shall argue, \ben\ has the potential of becoming
one of the ``calorimeters'' of choice in detecting non-thermal
hadronic processes during this early epoch of our Universe.  We also
investigate a similar ``tertiary'' production of~\bt.

\paragraph{Observations}

The elements Li, Be, and B are detected via atomic resonance lines in
the photospheres of MDSs.  The constancy of \lisv/H at lowest
metallicities $[\Fe/\Hyd]<-1.5$ is known as the ``Spite
plateau''~\cite{Spite:1982dd}, and is interpreted as representing the
primordial lithium component. Whereas numerous determinations
exist---with $\lisv/\Hyd$ in the range $(1- 2.5)\times 10^{-10}$---the
observational status of the isotopic ratio \lisx/\lisv\ remains
doubtful.
Though a plateau value of $\lisx/\lisv \simeq 0.05$ has been claimed
from detections in a number of halo stars~\cite{Asplund:2005yt}, this
claim has also been challenged. Since the \lisx\ absorption line is
not resolved spectroscopically from the one of \lisv, convective
motions may well mimic its presence~\cite{Cayrel:2007te}. Indeed, a
re-analysis~\cite{Steffen:2010pe} has diminished the number of such
($3\sigma$) detections to a mere number of two stars.

\begin{figure}
\includegraphics[width=\columnwidth]{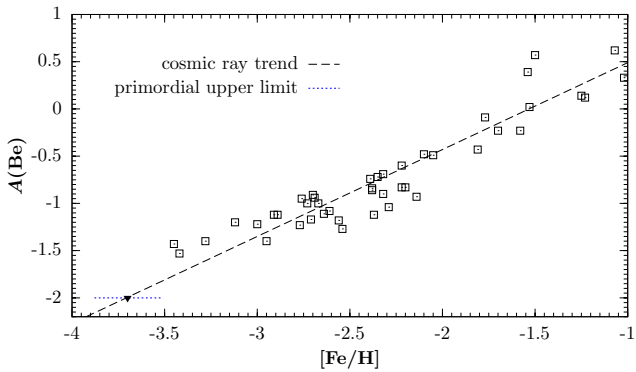}
\caption{\small Beryllium detections (squares) and fit of their Be-Fe
  trend taken from \cite{Rich:2009gj}. The triangle shows the upper
  limit from~\cite{Ito:2009uv}; $A(X)\equiv\log_{10}(X/\Hyd)+12$.}
\label{obs}
\end{figure}

In contrast to Li, Be and B show no indication of a plateau
structure. Instead, they exhibit a scaling with metallicity, which
points towards a pure cosmic ray origin.
The lowest observationally inferred B abundance then sets a limit on
its primordial value, $(\mathrm{B}/\Hyd)_p \lesssim
5\times 10^{-12}$~\cite{2010ApJ...713..458T}.  The determination of beryllium
is especially robust since \ben\ is the only stable
isotope. Figure~\ref{obs} shows a sample of \ben\ detections (taken
from \cite{Rich:2009gj}) as a function of metallicity. The lowest
point in black corresponds to the upper limit obtained recently
in~\cite{Ito:2009uv} and which we are going to use as an indicative
upper limit on primordial \ben:
\begin{align}
\label{eq:be9-limit}
  ({\ben}/{\Hyd})_p \lesssim 10^{-14} .
\end{align}
If instead only the detected values of \ben\ are used, a nominal
$3\sigma$ upper limit of $4.7 \times 10^{-14}$ can be deduced. In any
case, \ben\ is by far the rarest and the most constrained among the
light elements.

In addition to being firmly detected, there lies another virtue of Be
and B.  \lisx\ is fragile and may well have been destroyed in MDSs.
Such a destruction is indeed the essence of the ``astrophysical
solution'' to the \lisv\ problem, which posits a uniform stellar
depletion of \lisv\ by a factor 3 to 5 from its SBBN prediction
$\lisv/\Hyd = (5.24^{+0.71}_{-0.67})\times
10^{-10}$~\cite{Cyburt:2008zz}.
Any mechanism achieving this requires that the atmospheric material is
transported deep enough into the hotter stellar interiors where \lisv\
is destroyed.
If all three elements are subjected to the proton-burning reactions at
the same temperature, one obtains
\begin{align} 
\label{eq:depletion}
  {\lisv}/{\lisv_{p}} = \left( {\lisx}/{\lisx_p}\right)^a =
  \left( {\ben}/{\ben_p}\right)^b
\end{align}
with the exponents given by $a= \sigmavof{ \lisv+p} /
\sigmavof{\lisx+p }$ 
and $b= \sigmavof{ \lisv+p} / \sigmavof{\ben+p }$ (which makes the
estimate exponentially sensitive to temperature.)
Evaluating the cross sections for $\lisx(p,\alpha)\het$,
$\lisv(p,\alpha)\hef$, and $\ben(p,\alpha)\lisx$ as well as
$\ben(p,\deut)\beet$ at a fiducial temperature of $T=10^6\ \K$ one
finds $a\simeq 0.01$ and $b \simeq 3\times 10^5 $.  Irrespective of a
rather crude nature of (\ref{eq:depletion}), this estimate illustrates
one point: had \lisv\ been depleted by an $\Orderof{1}$-factor, \lisx\
could have been suppressed substantially, while \ben\ is to remain
unaffected
\footnote{For a critical discussion on estimates similar
  to~(\ref{eq:depletion}) in the context of concrete stellar depletion
  models see~\cite{2005ApJ...619..538R}.}.
Therefore, if stellar depletion is indeed a prevalent factor in the
reduction of \lisv, the use of \ben\ over \lisx\ for constraining
non-standard BBN scenarios may be preferred.  Table~\ref{t1}
summarizes the observational status of the rare light elements \lisx,
\ben, and $\mathrm{B} = \bt+\!\bel $, where we make a generous
allowance for the possible stellar depletion of primordial~\lisx.

\begin{table}
  \caption{\label{t1} 
    Rare light elements and (a) their 
    SBBN abundance prediction 
    (b) observationally inferred upper limit on the primordial abundance
    and (c) the temperature below which their BBN destruction by protons
    effectively stops; $\mathrm{B}=\bt+\!\bel$.}
  \begin{ruledtabular}
\begin{tabular}{cccc}
  element  & SBBN & 
primordial limit & $T_{\mathrm{f}}/\keV$  \\
  \lisx/\Hyd  & $\sim 10^{-14}$ 
&  $\lesssim \mathrm{few}\times 10^{-10}$ &  8  \\
  \ben/\Hyd   & $< 10^{-18}$  
& $\lesssim  10^{-14}$ &  10  \\
  $\mathrm{B}/\Hyd$    & $< 10^{-15}$ 
& $\lesssim 5\times 10^{-12}$ &  20 \\
\end{tabular}
\end{ruledtabular}
\end{table}

\paragraph{Conversion probability}

The key quantity that determines the efficiency of (\ref{eq:chain})
and ultimately the strength of \ben\ constraints is $P_{\mathrm{Be}}$,
that we define as the probability for an energetic $A=3$ nucleus
injected into the BBN plasma to produce \ben.  Clearly, the
chain~(\ref{eq:chain}) can be broken up into its sub-processes,
\begin{align}
\label{eq:P3to9}
  P_{\mathrm{Be}} & = \VEV{P_{3\to 6}}_{E_{3}^{\mathrm{in}}}\star
  \VEV{P_{6\to 9}}_{E_{6}^{\mathrm{in}}} \star \VEV{P_{9\to
      9}}_{E_{9}^{\mathrm{in}}} ,
\end{align}
where each individual probability $P_{A_1\to A_2}$ is averaged over
the distribution of $A_1$ injection energies $E_{A_1}^{\mathrm{in}}$.
It is easy to see that $P_{\mathrm{Be}}$ has to be a functional of the
initial energy distribution of the mass-3 element.  The stars indicate
that the quantities are not independent from each other but appear
under integral signs.

$P_{A_1\to A_2}$ with $A_1\neq A_2$ is found by integrating the
non-thermal interaction rate over the time that takes $A_1$ to slow
down below the $A_2$ formation threshold $E_{A_2}^{\mathrm{th}}$:
\begin{align}
\label{eq:P_A1toA2}
  P_{A_1\to A_2} =
  \int^{E_{A_1}^{\mathrm{in}}}_{\max\{E_{A_2}^{\mathrm{th}},E_{A_2}^{\mathrm{thm}}
    \}}dE_{A_1} \frac{n_{\hef} \sigma_{A_1\to
      A_2}(E_{A_1})}{|dE_{A_1}/dx|} .
\end{align}
Here, $n_{\hef}$ is the number density of ambient alpha particles and
$\sigma_{A_1\to A_2}$ is the cross section for the $A_2$ formation via
$A_1(\alpha,N)A_2$ with $N=n$ or $p$.  The energy loss rate
$dE_{A_1}/dx$ of $A_1$ is of great importance as it regulates the
overall efficiency of the \ben\ producing chain;
$E_{A_2}^{\mathrm{thm}}$ is the energy of $A_2$ once it is
thermalized. For the temperatures that are of most interest to us, $T
\leq 20$ keV, the thermalization rate is dominated by Coulomb stopping
on electrons and positrons in the plasma.

Equation (\ref{eq:P_A1toA2}) is to be averaged over the $A_1$ energy
distribution $f_{A_1}(E_{A_1}^{\mathrm{in}}) = \sigma_{A_1}^{-1}
d\sigma_{A_1}/dE_{A_1}^{\mathrm{in}} $; $\sigma_{ A_1}$~is the cross
section for producing $A_1$ secondaries from \hef. In general,
$f_{A_1}$ also depends on the energy of the primary which gives rise
to $A_1$. For the spallation products $\trit$ and $\het$ produced in
$N(\alpha,X)\trit/\het$ reactions, $f_{3}$ is largely independent of
the incident nucleon energy, and we shall use the fit to $f_3(E)$ as
the sum of exponential factors~\cite{Blinov:2006rb}; see
also~\cite{Cyburt:2009pg}.  When it comes to $A_1 = 6$, \ie\ \lisx\
and \hes, we assume a flat distribution of the differential cross
section with respect to the CM scattering angle.

When computing $\VEV{P_{6\to 9}}_{E_{6}^{\mathrm{in}}}$ one should
also account for the fact that $\hes$ may eventually decay
($\tau_{\hes} \simeq 0.8\,\sec$) before interacting
hadronically. This, however, has almost no influence
on~(\ref{eq:P3to9}). We find that even for temperatures $T$ as low as
$1\ \keV$ more than $90\%$ of \hes\ survives the initial degradation
from $E_{6}^{\mathrm{in}}$ to a kinetic energy of $ \sim 1\,\MeV$
below which the cross section for~(\ref{eq:rct-hes}) becomes too
small. Turning to the last factor in (\ref{eq:P3to9}), the probability
that a newly created \ben\ nucleus actually survives, we remark that
in our BBN code the proton burning of thermalized \ben\ (as well as of
\lisx\ and \hes) are already accounted for. Therefore, we identify
$\VEV{P_{9\to 9}}$ in (\ref{eq:P3to9}) with the ``in-flight''
destruction probability prior to thermalization. Finding $\VEV{P_{9\to
    9}}$ in an obvious modification of (\ref{eq:P_A1toA2}) we conclude
that the survival probability is always well above $90\%$ so that this
factor is of no importance.  On similar grounds, we can neglect any
potential ``in-flight'' $p$-destruction of $A=6$ secondaries.

\begin{figure}[t]
\includegraphics[width=\columnwidth]{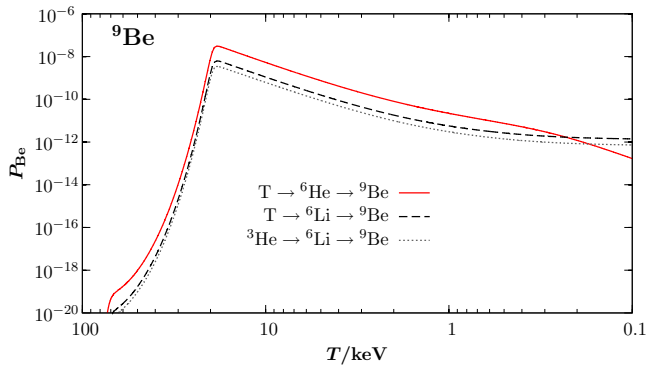}
\caption{\small Conversion probabilities $P_{\mathrm{Be}}$ as a
  function of photon temperature $T$ for the various channels
  depicted; thermal destruction of \ben\ is not included.}
\label{Fig:P}
\end{figure}

Figure~\ref{Fig:P} shows $P_{\mathrm{Be}}$ for the various channels as
labeled. The exponential drop in the number of electron-positron pairs 
leads to the sharp rise of $P_{\mathrm{Be}}$ with temperature, with the peak
at $T \simeq 20\ \keV$, followed by a slow decline caused by  
the logarithmic growth of the stopping power at lower temperatures.

We believe that our calculation of $P_{\mathrm{Be}}$ is accurate to
within a factor of two or three as it is largely based on experimental
data. When available, we made use of the (inverse) cross section data
collected in~\cite{Angulo:1999zz}. For the two nuclear reactions of
most interest, $\hes(\alpha,n)\ben$ and $\trit(\alpha,p)\hes$, we used
the measurements of~\cite{PhysRev.106.1252,Bass}
and~\cite{PhysRevC.17.1283}, respectively.  The former reaction shows
a large cross section above 100 mbn over a wide range of energies,
reaching a peak of 400 mbn at $E_6 \simeq 5\,\MeV$.  The latter yields
a maximum of $9\,\mbarn$ at $E_3 \simeq 20\,\MeV$ and was approximated
by a constant $S$-factor of $S\simeq 520\, \mbarn\,\MeV$ for its
inverse reaction $\hes(p,\alpha)\trit$.  Previous investigations of
secondary \lisx\ production have not accounted for the \hes\
channel. This is well justified, because the rate for
$\trit(\alpha,p)\hes$ reaction is considerably less than that of
$\trit(\alpha,n)\lisx$; the cross section for the latter as well as
for $\het(\alpha,p)\lisx$ we take from~\cite{Cyburt:2002uv}.

\paragraph{Constraints}

Having obtained $P_{\mathrm{Be}}$, we now compute the actual output
of~\ben\ in the spirit of previous works on non-equilibrium BBN.  The
induced non-thermal contribution to the Boltzmann equation is of the
form
$dn_{\ben}/dt|_{\mathrm{n.th.}} = n_{\hef} \Gamma_{4\to 3} \star
P_{\mathrm{Be}} $,
where $\Gamma_{4\to 3}$ is the destruction rate of \hef\ to $A=3$
final states.
Let $X$ decay with some hadronic branching fraction $B_h$ and denote
by $N_i(\e)$ the number of energetic nucleons per $X$-decay with
kinetic energy $\e$. Then, the spallation rate is given by
\begin{align}
\label{eq:gamma3to4}
  \Gamma_{4\to 3} = n_X \sum_{i=n,p}  \int_{E_{4,\mathrm{th}}}^{\e_{\mathrm{had}}/2}
  d\e\, N_i(\e)\, \sigma^{(i)}_{4\to 3}(\e) v_i(\e)
\end{align}
where $ \e_{\mathrm{had}}$ is the injected hadronic energy per
$X$-decay and $E_{4,\mathrm{th}}\sim 20\,\MeV$ is the threshold energy
in the $A=3$ producing reaction $p/n + \hef$ with cross
section~$\sigma_{4\to3}$.  We do not need to account for other
hadronic shower products such as long-lived mesons $\pi^{\pm},\,
K^{\pm}$ or $ K_L$. For cosmic times $t\gtrsim 100\,\sec$, they decay
before interacting hadronically---unless the $X$-abundance is in
excess of baryons~\cite{Pospelov:2010cw}.  We also find that
photodissociation channels such as $\hef(\gamma,N)\trit/\het$ and
$\lisv(\gamma,N)\hes/\lisx$ are not important for the production of
\ben.

The time evolution of $N_i $ is governed by the propagation
equation~\cite{Cyburt:2009pg}
\begin{align}
\label{eq:shower-prop}
\partial_t N_i = J_i - \Gamma_{i,\mathrm{sc}}
N_i -\partial_\e \left( \e\, \Gamma_{i,\mathrm{stop}} N_i \right) ,
\end{align}
where $\Gamma_{i,\mathrm{stop}} =- \e^{-1} dE_i/dt $ is the continuous
stopping rate, $ \Gamma_{i,\mathrm{sc}}$ is the total hadronic
scattering rate and $J_i$ is the sum of all source terms.  Up to
secondary contributions from elastic and inelastic down-scattering
processes, $J_i$ is given by $ J_i \simeq B_h Q_i(\e) / \tau_X$, where
$Q_i(\e)$ is the primary nucleon spectrum in the decay of~$X$ and
$\tau_X$ is the $X$-lifetime.  Annihilating $X$ are treated in
complete analogy with the replacement $ \tau_X^{-1} \to
n_X\sigmavof{\mathrm{ann}}$ where $\sigmavof{\mathrm{ann}}$ is the
annihilation cross section to hadrons.  Following
\cite{Kawasaki:2004qu,Jedamzik:2006xz} we simulate a generic hadronic
$X$-decay (and obtain $Q_i$) by using the
PYTHIA~\cite{Sjostrand:2006za} model for $e^{+}e^{-}$ annihilations to
hadrons.

\begin{figure}[t]
\includegraphics[width=\columnwidth]{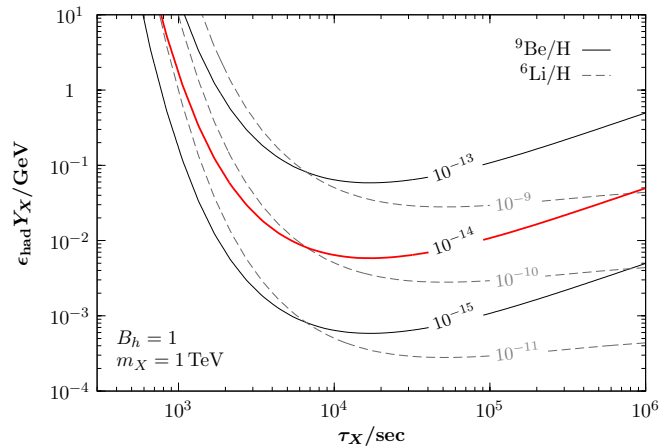}
\caption{\small Contours of constant \ben\ (solid lines) and \lisx\
  (dashed lines) as a function of $\tau_X$ and $\e_{\mathrm{had}}Y_X$;
  $B_h = 1$ and $m_X = 1\,\TeV$ is assumed.}
\label{Fig:C}
\end{figure}

From inspection of Table~\ref{t1} we expect a $\Orderof{1}$-surviving
fraction of non-thermally fused \lisx\ or \ben\ only for $T\lesssim
20\ \keV$. For these temperatures, neutrons are not stopped before
interacting hadronically. On the other hand, for protons with
$\epsilon \lesssim 1\ \GeV$ it is a reasonable approximation to assume
that the energy degradation term in (\ref{eq:shower-prop})
dominates. Then, employing a quasi-static equilibrium condition
$\partial_t N_i \simeq 0$ to (\ref{eq:shower-prop}) (a good
approximation for $\Gamma_{\mathrm{in}}\ll
\Gamma_{\mathrm{sc}},\Gamma_{\mathrm{stop}} $) one
finds~\cite{Cyburt:2009pg}
\begin{align}
\label{eq:qse}
   \frac{\e N_p(\e)}{ \Gamma_{p,\mathrm{in}} } & \simeq
  \frac{ 1 }{ \Gamma_{p,\mathrm{stop}}} \int_{\e}
  d\e'\, Q_p(\e'), \quad
  \frac{ \e N_n(\e) }{\Gamma_{n,\mathrm{in}}}\simeq \frac{\e Q_n(\e)
  }{\Gamma_{n,\mathrm{sc}}} .
\end{align}
Using (\ref{eq:qse}) to obtain $\Gamma_{4\to3}$
in~(\ref{eq:gamma3to4}), we solve the BBN network of Boltzmann
equations for various combinations of $\tau_X$ and $X$-abundance
$Y_X$. The latter is normalized on the number density of baryons,
$Y_X\equiv n_X/n_b$.

Figure~\ref{Fig:C} shows contour lines of constant $\ben/\Hyd$ and
$\lisx/\Hyd$ as a function of $\tau_X$ and
$\e_{\mathrm{had}}Y_{X}$. We have assumed $B_h=1$ so that
$\e_{\mathrm{had}}=m_X=1\,\TeV$. The (solid) \ben\ contours, when
compared with the ones for \lisx\ (dashed), illustrate that \ben\ is
more sensitive to energy injection at earlier times, reaching its
maximal abundance for $\tau_X\simeq 1.5\times 10^4\,\sec$.  For larger
lifetimes, as the primordial gas keeps rarefying, it becomes
increasingly difficult to produce \ben\ from $A=6$. For $\tau_X\gtrsim
10^6\,\sec$ the hadronic process looses further ground as neutrons
decay before interacting strongly. At optimal $\tau_X$, the current
limits on primordial \ben\ probe the energy injection at a level of 10
MeV per nucleon, or $10^{-12}$ GeV per entropy.  More elaborate
treatments along the lines of
\cite{Kawasaki:2004qu,Jedamzik:2006xz,Cyburt:2009pg} will certainly
help to improve on the estimate of produced \ben\ per unit injected
energy. It is important to note, however, that the ratio \ben/\lisx,
and thus their relative importance as constraints will not be affected
by that. Another (competitive) constraint on hadronic energy injection
is due to overproduction of $\deut/\Hyd$; for a comprehensive survey
on general energy injection confer the works cited above.

Can a chain of reactions similar to~(\ref{eq:chain}) lead to the
fusion of \bt?  One way to increment $A$ within the
chain~(\ref{eq:chain}) by one unit is to replace one of its photonless
recoil reactions with a radiative $\alpha$-capture reaction, such as
$\lisx(\alpha,\gamma)\bt$ or $\het(\alpha,\gamma)\bes$ followed by
$\bes(\alpha,p)\bt$.
The associated cross sections are, however, only in the $\mu\barn$
range, resulting in a \bt\ conversion probability that is typically
three orders of magnitude below the one of \ben.
A more prospective avenue is to start off a sequence of
transformations with energetic \hef\ rather than with \trit\ or
\het. The former can be obtained in interesting quantities via elastic
scatterings of shower-nucleons on ambient $\alpha$ particles. This
gives way to the formation of energetic $A=7$ from which $\bt$ (and
$\beten$) can be fused. By explicitly modeling this path we find,
however, that the $A=4$ to $A=10$ conversion probability
$P_{\mathrm{B}}$ typically remains below $10^{-10}$ so that \bt\ is
far from reaching observationally interesting levels.

\paragraph{Conclusions}

In this letter we show that the production of ``secondary''
\lisx~\cite{Dimopoulos:1987fz} in non-equilibrium nucleo\-synthesis
scenarios can be followed up by a ``tertiary'' formation of \ben/\Hyd\ in
observationally relevant quantities~\mbox{$\gtrsim 10^{-14}$}.  The
ensuing constraints on models of new physics are similar to those
imposed by \lisx, but do not suffer from the uncertainty associated
with stellar depletion.  Future observations of stars at lowest
metallicities may strengthen this constraint even further, which sets
\ben\ apart from other light elements where such improvements appear
unlikely.

\end{document}